\documentclass{elsart3}

\usepackage{graphicx}

\usepackage{amssymb}

\begin{document}

\begin{frontmatter}



\title{Heavy fermion $ d $ wave superconductivity: a X-boson approach}
\thanks[CNPq]{We acknowledge 
the financial support from the Rio de Janeiro
State Research Foundation (FAPERJ)
and National Research Council (CNPq).}


\author[IfUFF]{Lizardo H. C. M. Nunes}
\author[IfUFF]{M. S. Figueira}
\author[IfUFF]{E. V. L. de Melllo}

\address[IfUFF]
{Instituto de F\'{\i}sica da 
Universidade Federal Fluminense\\
Av. Litor\^{a}nea s/n, 24210-340 Niter\'oi, Rio de Janeiro, Brasil.
C.P.100.093}

\begin{abstract}
From an extension of the periodic 
Anderson model (PAM) in the $ U=\infty $ limit
taking into account the effect of a nearest
neighbor attractive interaction between $ f $-electrons,
we compare the obtained superconducting phase diagram
of a two dimensional d-wave superconductor
with the results obtained for an isotropic s-wave superconductor
employing the X-boson method.
\end{abstract}

\end{frontmatter}

\section{Introduction}\label{Int}

Recently, we have used the X-boson method \cite{X-boson}
to study the superconducting phase diagram of the 
PAM in the $ U \rightarrow \infty $,
where $ U $ is the on-site 
Coulomb repulsion, in the presence of
an attractive interaction between
the localized $ f $-electrons, 
\begin{eqnarray} \label{model}
H  & = & \sum_{ {\bf k}, \sigma } {\epsilon}_{ {\bf k}, \sigma}
c^{\dagger}_{ {\bf k}, \sigma} c_{ {\bf k}, \sigma} + \sum_{ {\bf
k}, \sigma} {E}_{f} X_{{\bf k},\sigma \sigma }
\nonumber \\
& &
+ \sum_{{\bf k}, \sigma } V (X_{\bf
{k},0\sigma}^{\dagger}c_{{\bf k},\sigma} +  c^{\dagger}_{\bf {k},
\sigma}X_{\bf {k}, 0\sigma } )
\nonumber \\
& &
+ \sum_{ {\bf k}, {\bf k'} } W_{ {\bf k}, {\bf k'} }
b^{\dagger}_{ {\bf k} } b_{ {\bf k'} } \, ,
\end{eqnarray}
where we only considered 
site independent local energies
${ E }_{f,j,\sigma}={ E }_{f,\sigma}$
and a constant hybridization
$ V = V_{j,\sigma,\mathbf{k}} $.
The operator
$
b^{\dagger}_{{\bf k } } = X_{\bf {k},0\sigma}^{\dagger} X_{-\bf
{k},0\overline{\sigma}}^{\dagger} 
$,
creates Cooper pairs, where
$ X_{ {\bf k } ,\sigma \sigma'} $
is the Hubbard operator.
We have shown \cite{Nunes} 
that the existence of superconductivity 
is constrained to a region where the 
$ f $-band densities of states
$ \rho_{f}( \omega ) $ at $ \omega = \mu $ is sufficiently high.
A high $ \rho_{f}( \mu )$ usually 
implicates higher values of $ T_{c} $,
what suggests that
the Kondo behavior of the system favors superconductivity
in the X-boson approach.
Nevertheless, as charge carriers are added 
to the system
the superconductivity was found both for
configurations where the system presented intermediate valence (IV) and heavy fermion (HF) behavior
and we have recovered the three
characteristic regimes of the PAM:
Kondo, IV and  magnetic. This behavior which cannot be found for the
same model by the slave-boson treatment \cite{portugueses},
since it breaks down 
when the $ f $-occupation number $ N_{ f } \rightarrow 1 $.

Despite the fact that the model was primarily designed
to the study of the heavy fermion compounds,
theoretical descriptions of the 
superconducting phases based on two-band
models have application for a large variety of systems and
the model could fit into the description of
the high temperature superconductors
compounds (HTSC) \cite{HTSCModel}
or into the new two-band superconductor
MgB$_{2} $. Meanwhile, unconventional superconductors
may exhibit different pair symmetries.
For instance, 
HTSC cuprates appears to present 
$ d_{ x^{2} - y^{2} } $ superconductivity
for some compounds and
$ s $-wave for others
while for the HF compounds 
the crystallographic anisotropies and
strong spin-orbit coupling 
make the task of determining the 
pair symmetry very difficult,
athough $ d $-wave model is still one
of the leading candidates 
to describe superconductivity in UPt$_{3} $
\cite{Miyake86}.

In this paper we employ the X-boson approach
to compare the results for the 
superconducting phase diagram of an isotropic $ s $-wave superconductor with the results provided by a 
$ d_{ x^{2} - y^{2} } $ superconductor.

\section{The Method}\label{theMethod}

The X-boson approach consists of introducing
the variational parameter
$ R=1-\sum_{\sigma}<X_{\sigma\sigma}>$,
which modifies the Green's functions (GF) so that
it minimizes an adequate thermodynamic potential while being
forced to satisfy the ``completeness'' relation
$ n_{0}+n_{\sigma}+n_{\overline{\sigma}}=1 \label{Eq.2} $,
where $ n_{j,a}=<X_{j,aa}> $.
To this purpose we added to the model
the product of each ``completeness'' relation into a Lagrange
multiplier $\Lambda_{j}$, and use this new Hamiltonian to
generate the functional that shall be minimized by employing
the Lagrange's method. 

At the critical temperature 
the Green's functions yields to the previous
result found in the chain approximation
of the PAM, 
\cite{Hewson}
but now with the localized
energy levels $ E_{ f } $ and $ D_{\sigma} = R + n_{\sigma } $
renormalized,
and we solve the
mean-field
superconducting gap equation \cite{Nunes}
at $ T = T_{ c } $
constrained to 
the ``completeness'' relation above. 
For the  two dimensional 
$ d_{ x^{2} - y^{2} } $ superconductor
we use the tight-binding band dispersion
$ \epsilon({\bf k } ) = -2t \sum_{i=x,y} cos( k_{i} ) $
and an attractive interaction with
a $ cos( k_{x} ) - cos( k_{y} ) $ dependence in
$ {\bf k } $-space leads to the 
same superconducting gap symmetry.
On the other hand,
the superconducting phase diagram
for the $ s $-wave isotropic gap is
obtained using a constant conduction
density of states,
$ \rho( \epsilon_{\bf k }) = 1 / 2D $,
only defined for the interval
$  -D \le \epsilon_{\bf k } - \mu \le D $.
We make $ E_{ f } = -0.15 $, $ V =0.3 $ and 
the superconducting interaction
$ W = -0.20 $ in units of the half-bandwidth $ D $ 
or the hopping parameter $ t $. 
Finally, the total electron number 
$ N_{ t } = N_{ f } + N_{ c } $,
is kept constant and the
chemical potential is calculated
self-consistently  while 
the electrons are allowed   
to transfer between bands.

\section{Results}\label{results}
\begin{figure}[htb]
\centerline
{
\includegraphics[
angle=90,
width=0.43\textwidth]
{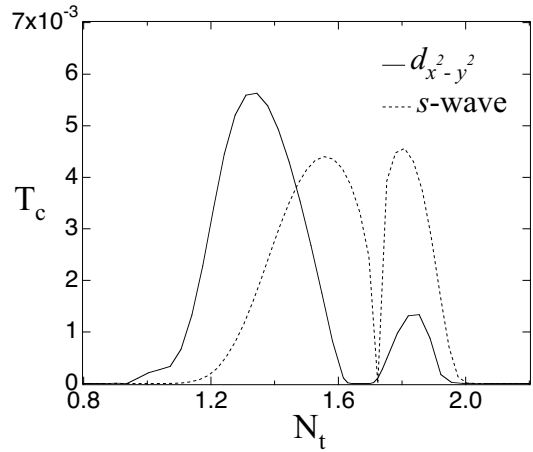}
}
\caption
{
The superconducting
phase diagram for  a
2-$d $  $d_{x^{ 2 } - y^{ 2 } } $ superconductor
(filled line) and an 
isotropic $ s $-wave superconductor (dashed line).
$ E_{ f } = -0.15 $, $ V =0.3 $ and 
$ W = -0.20 $.
} 
\label{FigTcxNt_V03}
\end{figure}
In Fig. \ref{FigTcxNt_V03} the superconducting
phase diagrams for a 
2-$d $  $d_{x^{ 2 } - y^{ 2 } } $ superconductor
and an isotropic $ s $-wave superconductor
are shown. 
As previously obtained \cite{Nunes}
the superconductivity 
is constrained to a region where 
$ \rho_{f}( \mu )$ is high and our results
show that $ d $-wave pairing exhibits the highest $ T_{ c } $
up to occupations of about $ N_{ t } \approx 1.4 $,
which is in the vicinity of the Kondo regime.
For  higher occupations,
both systems present 
a superconductor-insulator transition
when the chemical potential $\mu $
crosses the hybridization gap, 
what means that $\mu $ lies in the region 
between the peaks of the density of states
and the system presents a IV behavior.
Notice that for even more higher occupations,
$ \mu $ reaches the upper band,
what cannot be obtained by the slave-boson method,
since it breaks down when $ N_{f } \rightarrow 1 $.
For $ N_{t } \gtrsim 1.4 $
a crossover occurs into a
regime where the isotropic $  s $-wave pairing
becomes the most stable 
for the parameters considered.



\end{document}